# REC language is a live on IBM1130 simulator

**This work is archaeological reconstruction of REC/A language on IBM1130 Simulator from Computer History Simulation Project**

Ignacio Vega-Páez[1], José Angel Ortega[2] and Georgina G. Pulido[3]

ivegap0800@ipn.mx, oeha430210@hotmail.com and gpulido@att.net.mx

IBP-TR2009-04

Apr 2009, México, D.F.

**ABSTRACT**

REC (Regular Expression Compiler) is a concise programming language development in mayor Mexican Universities at end of 60's which allows students to write programs without knowledge of the complicated syntax of languages like FORTRAN and ALGOL. The language is recursive and contains only four elements for control. This paper describes use of the interpreter of REC written in FORTRAN on IBM1130 Simulator from "Computer History Simulation" Project [2008 Vega].

**Terms Key:** REC (Regular Expression Compiler), Programming language, Archeology software.

This work is archaeological reconstruction of language REC/A on IBM1130 Simulator form Computer History Simulation Project (http://simh.trailing-edge.com/) used for 1130 fans in www.ibm1130.org, the interpreter REC Fortran was writing for Gerardo Cisneros at the beginning of computing in Mexico, which marks a milestone in software development in Mexico, so it is important to bring to life this version of REC/A. A formal definition of REC language was published by Harold V. McIntosh in AIM-149 of MIT Artificial Intelligence Group and too "Acta Mexicana de Ciencia y Tecnología" of IPN México City. see [68a McIntosh] & [68b McIntosh] respectively, and for detail of REC in FORTRAN published by Gerardo Cisneros see [70 Cisneros] and samples of language REC in "Lectures notes on programming FORTRAN" from ESFM-IPN see [79 Rojas] .

---

[1] Postgrade Studies and Research Section of Mechanic and Electric Engineering Scholl, of the Polytechnic National Institute (SEPI ESIME IPN) and the Technology Information direction of the enterprise International Business Partners consulting (IBP-Consulting)
[2] Post grade Studies and Research Section of Mechanic and Electric Engineering Scholl, of the Polytechnic National Institute (SEPI ESIME IPN)
[3] Basic Sciences and Engineering Department of Metropolitan Autonomous University (UAM AZC), Unit Azcapozalco.

The next steps are use language REC/A on IBM1130 simulator.

1. To use the emulator on Windows, download `ibm1130software.zip` from www.ibm1130.org or http://simh.trailing-edge.com/ and unzip it into a working directory, say `\ibm1130simh` This directory will contain the Windows executables and the sample job files

2. Start the emulator by typing the command **`ibm1130`** show two windows: 1130 panel and simulator console, it is ready for JOB.

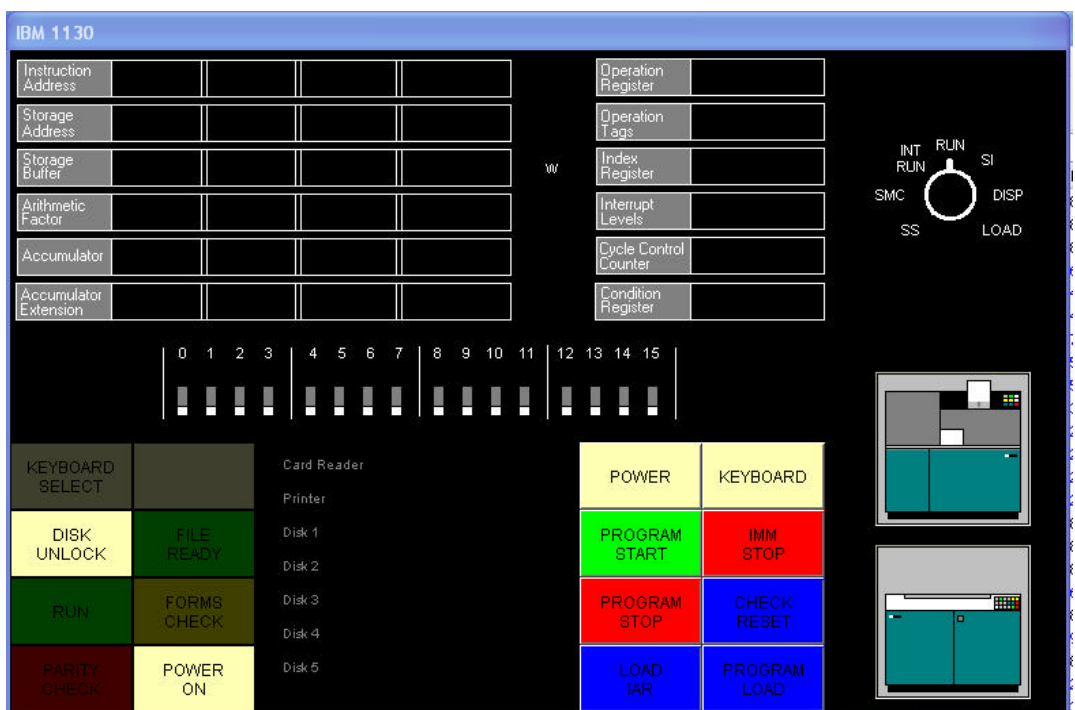

Panel of the IBM 1130

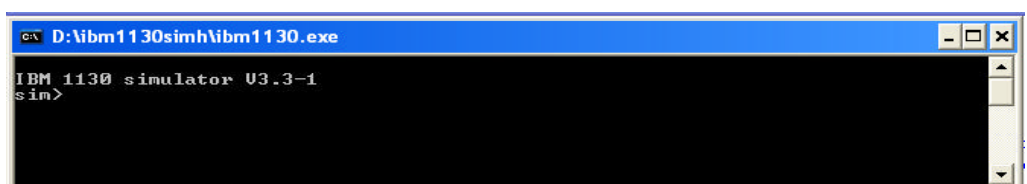

Simulator console

3. Compiling routines of REC and allocate data tables into Disk system.

    a. For compiling routines of REC, use command

    **DO JOB REC_C**

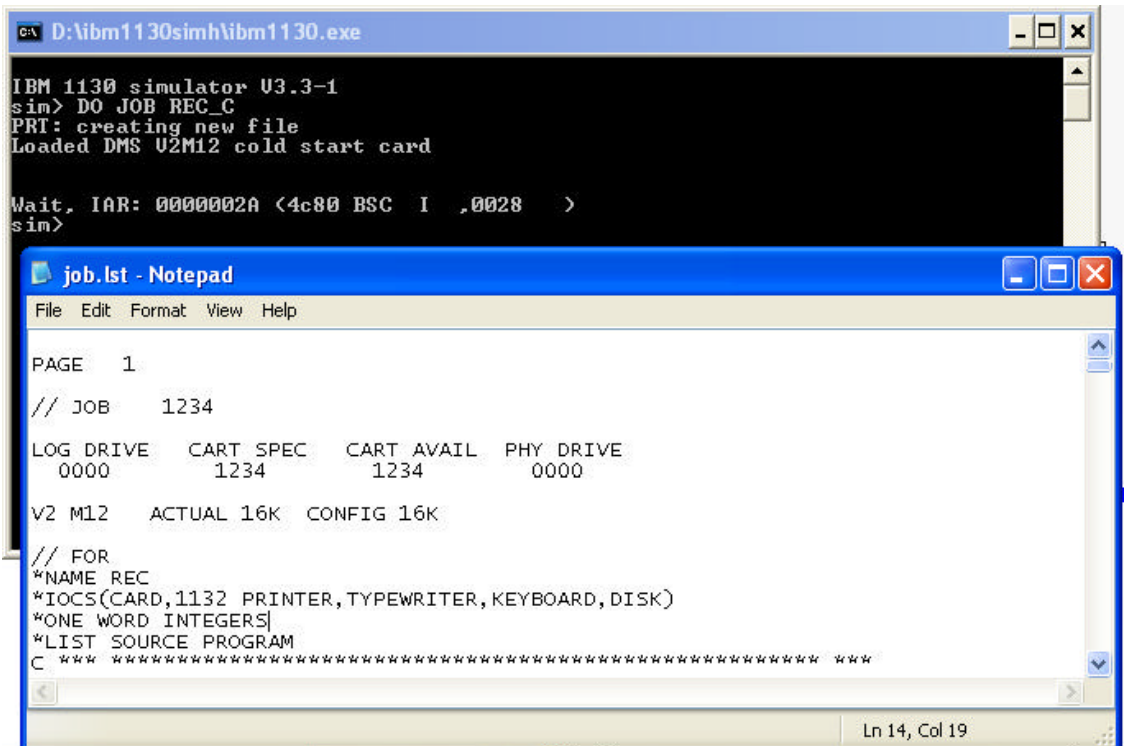
Output results in listing file **job.lst**

b. Data tables of execution time for REC, use commands

   `DO JOB RECDO` and `DO JOB RECER`

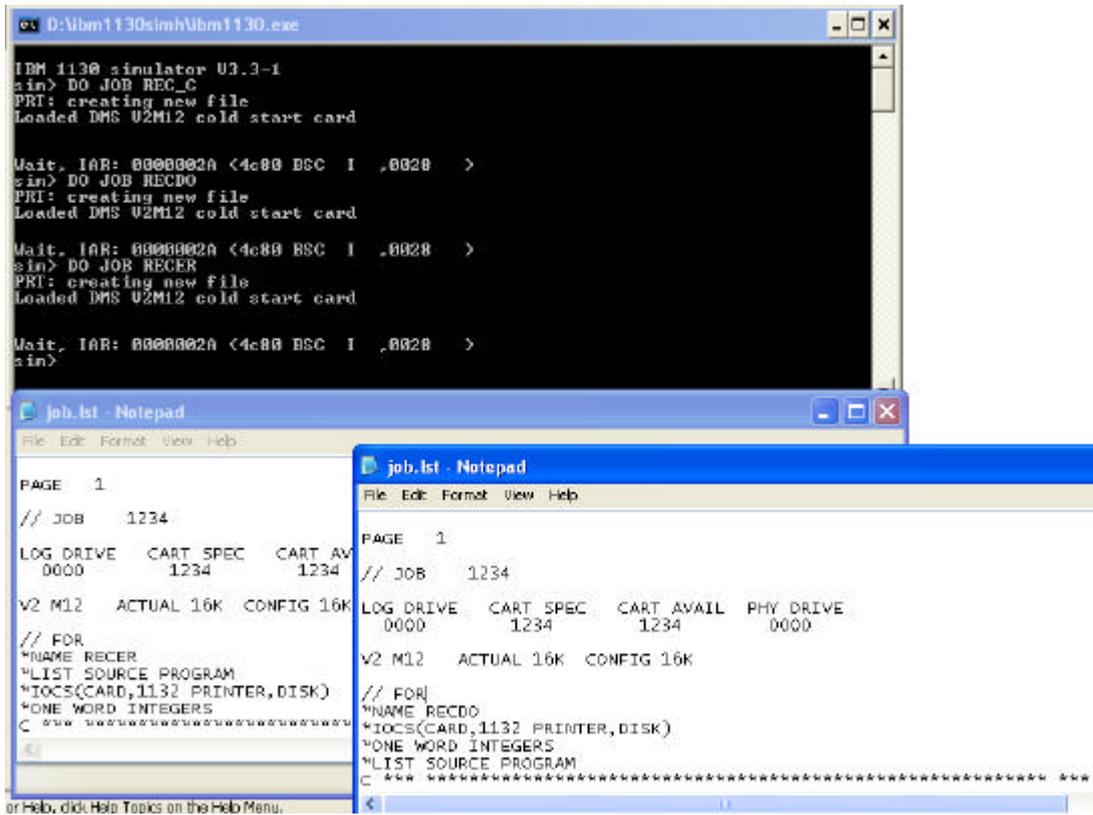

Output for each command it is in listing file **job.lst**

The language **REC** is ready for use.

4. For executing samples, use command:

```
DO JOB REC
```

```
PAGE    1

// JOB    1234

LOG DRIVE    CART SPEC    CART AVAIL    PHY DRIVE
  0000         1234          1234          0000

V2 M12    ACTUAL 16K    CONFIG 16K
0
PAGE    1

// JOB REC

LOG DRIVE    CART SPEC    CART AVAIL    PHY DRIVE
  0000         1234          1234          0000

V2 M12    ACTUAL 16K    CONFIG 16K

// * 690017 ESCUELA SUPERIOR DE FISICA Y MATEMATICAS

// XEQ REC      2

*LOCALREC,RECCA,RECCC,RECCO,RECKO,RECMD,RECNP,RECQU

*FILES(1,RECAT),(2,RECEM)

C DAMPED OSILLATIONS (Y=SIN(3*X)*EXP(-0.3*X)
* (
'/ ' S1L(F10'/3'*'SF1'/-0.3'*E*0'/1.0'&(N''*','' ''/-0.04'&.)LXF1'/0.15'&S1L   $50$.,),)
 0.00000E 00   0.00000E 00
 1.50000E-01   4.15826E-01
 3.00000E-01   7.15906E-01
 4.50000E-01   8.52504E-01
 6.00000E-01   8.13425E-01
 7.50000E-01   6.21304E-01
 9.00000E-01   3.26253E-01
 1.05000E 00  -6.13488E-03
 1.20000E 00  -3.08735E-01
 1.35000E 00  -5.25927E-01
 1.50000E 00  -6.23300E-01
 1.65000E 00  -5.92444E-01
 1.80000E 00  -4.50328E-01
 1.95000E 00  -2.33855E-01
 2.10000E 00   8.95249E-03
 2.25000E 00   2.29140E-01
 2.40000E 00   3.86318E-01
 2.55000E 00   4.55686E-01
 2.70000E 00   4.31465E-01
 2.85000E 00   3.26366E-01
 3.00000E 00   1.67559E-01
 3.15000E 00  -9.70736E-03
 3.29999E 00  -1.70005E-01
 3.44999E 00  -2.83736E-01
 3.59999E 00  -3.33121E-01
 3.74999E 00  -3.14203E-01
 3.89999E 00  -2.36499E-01
 4.04999E 00  -1.20008E-01
 4.19999E 00   9.53147E-03
 4.34999E 00   1.26089E-01
 4.49999E 00   2.08370E-01
 4.64999E 00   2.43504E-01
 4.79999E 00   2.28793E-01
 4.94999E 00   1.71356E-01
 5.09999E 00   8.59142E-02
 5.24999E 00  -8.69345E-03
 5.39999E 00  -9.34868E-02
 5.54999E 00  -1.53006E-01
```

```
IBM 1130 simulator V3.3-1
sin> DO JOB REC_C
PRT: creating new file
Loaded DMS V2M12 cold start card

Wait, IAR: 0000002A (4c80 BSC  I  ,0028   >
sin> DO JOB RECDO
PRT: creating new file
Loaded DMS V2M12 cold start card

Wait, IAR: 0000002A (4c80 BSC  I  ,0028   >
sin> DO JOB RECER
PRT: creating new file
Loaded DMS V2M12 cold start card

Wait, IAR: 0000002A (4c80 BSC  I  ,0028   >
sin> DO JOB REC
PRT: creating new file
Loaded DMS V2M12 cold start card

Wait, IAR: 0000002A (4c80 BSC  I  ,0028   >
sin>
```

Output for samples REC

Next code is REC code of samples from several papers or thesis on REC language.

```
            // JOB REC
            // * 690017 ESCUELA SUPERIOR DE FISICA Y MATEMATICAS
            // XEQ REC        2
            *LOCALREC,RECCA,RECCC,RECCO,RECKO,RECMO,RECNP,RECQU
            *FILES(1,RECAT),(2,RECEM)

            C DAMPED OSILLATIONS (Y=SIN(3*X)*EXP(-0.3*X)
            * ('/ 'S1L(F1O'/3'*'SF1'/-0.3'*E*O'/1.0'&(N''*','' ''/-0.04'&.)LXF1'/0.15'&S1L
            $50$.,),)

            C THE EXPRESSION F(X,Y)=(X*X&Y*Y)**5-(8*(X*X-Y*Y)*X*Y)**2
            C IS EVALUATED 50*74 TIMES. THE CONDITION F(X,Y)=0 IS THE
            C BOUNDARY CURVE R = 2*SIN(4*THETA). AN * IS PUT OUT IF
            C F(X,Y) IS NEGATIVE AND A BLANK IF IT IS POSITIVE
            * ('/-2'S0L($50$'/-2'S1L($74$F1P*F0P*&PPPP****F1P*F0P*-
            F1*F0*'/8'*P*-(N''*','' ',)LF1'/0.054'&S1L.)XF0'/0.08'&S0L.,),)

            C FACTORIAL DE NUMERO
            * N'R
            (N,0L'/1',P'/1'-'*RECURSION' 'R*,)'RL
            ('/ 'S0L($10$F0'/1'&S0 O 'R O X.,),) L

            C SIMPSON INTEGRATION
            C   4*I(0,1)(1/1&X**2)
            *
            (F6 P F1 P * & / ,)Y
            (IOIOXS1-IOXS3'/2'*/S4'/3'/S5L'/1'S6'/'
            (F3F6-S3N,LY&F1F4&S1LY'/4'*&F1F4&S1LY&X.)LF5*,)'R
            ('R'/4'*''PI='OX,)
             '/1'
             '/0'
             '/40'

            C ****************************************************************
            C OTROS EJEMPLOS
            C ****************************************************************
            C EJEMPLOS DE LAS NOTAS DE RAUL ROJAS
            C ****************************************************************
            C LECTURA Y ESCRITURA DE UN NUMERO
            C SUMA DOS CONSTANTES
            * ('/2''/2'*O,)

            C SUMA DOS NUMEROS DESDE LA UNIDAD DE TARJETAS
            * (I''LA SUMA DE'OI''+'O''='+O,)
             '/10'
             '/2'

            * (I''1ero'OI'' 2do'OX-0''NUMEROS =s',N''EL 2do ES MAYOR','EL 1ro ES MAYOR',)
             '/10'
             '/2'

            * N'F
             (0L'/1',P'/1'-'F*,)'FL
             (I ''EL FACTORIAL DE 'O '' ES:' 'FOX,)
            '/4'

            C GRAFICA SEL SENO
            * ('/0.0628'S0L'/0'S1L($100$F1'S'/1'+('/0.04'-N''*','' '.)XLF1F0+S1.,),)

            C ****************************************************************
            C EJEMPLOS DE LA TESIS DE (REC;;) DE HECTOR SALDAÑA
            C ****************************************************************
            C EJEMPLO 1, LECTURA Y ESCRITURA DE UN NUMERO
            *(I O ,)
             '/2.4'
            C EJEMPLO 2, LECTURA DE DOS NUMEROS SE HACE LA SUMA Y SE IMPRIME
```

```
*(I I + O )
 '/5.4'
 '/3.2'
C EJEMPLO 3, UNA TABLA DE VALORES DE LOS 10 PRIMEROS DIGITOS Y SUS RAIZ CUADRADA
C           EL OPERADOR D IGULA A P (DUPLICA EL TOPE DE LA PILA), J IGUAL A L (ELIMINA
EL TOPE DE LA PILA
*('/0' ($10$ P ''     ' O ''    ' Q O L X '/1' + . ) )
```

The system REC files are:

- REC_C, contains REC compiler writing in FORTRAN.

- RECDO.JOB and RECER.JOB contains data setup for load into IBM 1130 system disk, RECDO load tables for compilation and execution y RECER error messages of system REC.

- REC.JOB contains REC code samples.

The files are available from author ivegap0800@ipn.mx